\documentclass[preprint,showpacs,preprintnumbers,amsmath,amssymb]{revtex4}
\usepackage{graphicx}
\usepackage{dcolumn}
\usepackage{bm}
\begin{document}
\title{High frequency longitudinal and transverse dynamics in water}
\author{
\small{ E. Pontecorvo$^{1,4}$,
        M. Krisch$^{2}$,
        A. Cunsolo$^{3}$,
        G. Monaco$^{2}$,
        A. Mermet$^{2}$,
        R. Verbeni$^{2}$,
        F. Sette$^{2}$,
        and G. Ruocco$^{1,4}$
}
        }
\address{
\small{
    $^{1}$Dipartimento di Fisica and INFM, Universit\'a di Roma ``La Sapienza'', I-00185, Roma, Italy.\\
} \small{
    $^{2}$European Synchrotron Radiation Facility, B.P. 220, F-38043 Grenoble, France.\\
} \small{
    $^{3}$Institut Laue-Langevin, BP 156, F-38043 Grenoble, France.\\
} \small{
    $^{4}$CRS SOFT, INFM, Universit\'a di Roma ``La Sapienza'', I-00185, Roma, Italy.
}
    }
\date{\today}
\begin{abstract}
High-resolution, inelastic x-ray scattering measurements of the
dynamic structure factor $S(Q,\omega)$ of liquid water have been
performed for wave vectors Q between 4 and 30 nm$^{-1}$  in
distinctly different thermodynamic conditions (T= 263 - 420 K ;
at, or close to, ambient pressure and at P = 2 kbar). In agreement
with previous inelastic x-ray and neutron studies, the presence of
two inelastic contributions (one dispersing with $Q$ and the other
almost non-dispersive) is confirmed. The study of their
temperature- and $Q$-dependence provides strong support for a
dynamics of liquid water controlled by the structural relaxation
process. A viscoelastic analysis of the $Q$-dispersing mode,
associated with the longitudinal dynamics, reveals that the sound
velocity undergoes the complete transition from the adiabatic
sound velocity ($c_0$) (viscous limit) to the infinite frequency
sound velocity ($c_\infty$) (elastic limit). On decreasing $Q$, as
the transition regime is approached from the elastic side, we
observe a decrease of the intensity of the second, weakly
dispersing feature, which completely disappears when the viscous
regime is reached. These findings unambiguously identify the
second excitation to be a signature of the transverse dynamics
with a longitudinal symmetry component, which becomes visible in
the $S(Q,\omega)$ as soon as the purely viscous regime is left.
\end{abstract}
\maketitle
\section{Introduction}
Water occupies a prominent role in natural sciences, partly due to
its relative abundance and central role for the existence of life
on earth, but as well due to its many unusual properties, which -
despite intensive research efforts - still defy today a complete
understanding \cite{1,2}. It is believed that the peculiar
physico-chemical behavior of water arises from its local structure
which is characterized by an open three-dimensional hydrogen-bond
network of water molecules with an almost perfect tetrahedral
arrangement and a well-structured second coordination shell. The
hydrogen bond is responsible for this highly ordered local
structure, and it is therefore clear that the processes of H-bond
breaking and formation play a central role in determining the
dynamical and thermodynamical properties of liquid water. The
lifetime of the H-bond is, at ambient conditions, in the
picosecond range. Thus, the study of the water dynamics in the
terahertz frequency range is of particular interest.
Traditionally, this is done by inelastic scattering techniques
using neutrons (INS) or x-rays (IXS), or by molecular dynamics
(MD) studies. More recently, the high-frequency dynamics has been
studied experimentally by time-domain spectroscopy \cite{3,torre}.
The key quantity in most of these studies is the so-called
dynamical structure factor $S(Q,\omega)$ ($\hbar Q$ and
$\hbar\omega=E$ denote the momentum and energy transfer,
respectively), which is the time and space Fourier transform of
the atomic density-density pair correlation function \cite{4}.
Starting from the pioneering work by Bosi et al. \cite{6} and by
Texeira et al. \cite{5}, several neutron experiments
\cite{7,8,sacchnew} and x-ray studies \cite{9a,9b,25,9c,10,9d,11}
have been devoted to the study of the THz-frequency dynamics in
liquid water.

The most striking result of these INS, IXS and MD studies is the
existence of a particularly large dispersion effect in the sound
velocity as a function of frequency. More specifically, at T = 277
K and ambient pressure, the longitudinal sound velocity increases
from its hydrodynamic value $c_0 = 1500 m/s$ to $c_\infty \approx
3200 m/s$. This transition from the hydrodynamic, or
zero-frequency, to the infinite-frequency sound regime is directly
observed at Q = 2 nm$^{-1}$ and energy E = 3 meV. This phenomenon
bears strong resemblance with that observed in glass-forming
liquids, where the sound velocity dispersion is due to the
presence of a structural (or $\alpha$) relaxation process. If
$\tau_\alpha$ is the characteristic time of this process, the
system has a solid-like elastic behavior for modes with frequency
$\Omega(Q)$ satisfying $\Omega (Q) \tau_\alpha>> 1$ and a viscous
one for frequencies such that $\Omega (Q) \tau_\alpha<< 1$ (here
$\Omega (Q) =c(Q) Q$ is the excitation frequency). A detailed
viscoelastic analysis of the dynamic structure factor demonstrated
that also in water the transition from low- to high-frequency
sound velocity is actually driven by the structural relaxation
process \cite{10}.

Temperature-dependent IXS studies, in which the density was kept
approximately constant to 1 g/cm$^{3}$ by changing the pressure,
revealed that, as long as $\tau_\alpha$ remains in the picosecond
region (i.~e. in the high temperature region), it follows an
Arrhenius behavior with an activation energy comparable to the
hydrogen bond energy. This result provided a link between the
relaxation process and the hydrogen bond network, and offered an
explanation for the microscopic origin of the structural
relaxation process in water. On a time scale short with respect to
the lifetime of these H-bonded local structures the collective
dynamics is very similar to that of the solid state, i.e. ice
\cite{9b}. In the opposite limit, the local structures have
sufficient time to relax, and therefore show a typical liquid
behavior. In the intermediate region, for $\Omega (Q)
\tau_\alpha\approx 1$, the dynamics of the density fluctuations is
strongly coupled with the making and breaking of the hydrogen bond
network.

A second important result, arising from the whole body of INS and
IXS studies of liquid water, is the existence of a second
excitation in the dynamic structure factor. This second excitation
has a (almost) $Q$-independent energy, and becomes visible only at
$Q$ larger than $\approx$4 nm$^{-1}$.  This second feature,
observed in the $S(Q,\omega)$ of ambient condition liquid water by
INS \cite{6,7,8,sacchnew} and IXS \cite{9b,25,9d}, has been
attributed to transverse-like dynamics on the basis of a MD study
\cite{22} where the comparison between longitudinal and transverse
current spectra was performed. The existence of the second feature
and its transverse origin can be explained within the same
viscoelastic framework which has been successfully used to
describe the longitudinal dynamics. Indeed, the transverse
dynamics, known to be non-propagating in liquids, can be actually
supported by the liquid structure as soon as the viscous regime is
left and the response becomes solid-like. The "transverse"
excitation acquires finite intensity in the $S(Q,\omega)$ - which
is only sensitive to the longitudinal molecular motion - because
of the lack of order in the structure, leading to a mixing of
modes with longitudinal and transverse symmetry \cite{22}.

To summarize, the high frequency dynamics of liquid water seems to
be controlled by a relaxation process -whose microscopic dynamics
is related to the making and breaking of the hydrogen bond
network- with a characteristic time $\tau_\alpha$. All the modes
with frequencies such that $\Omega (Q) \tau_\alpha << 1$ see a
relaxing local structure (viscous regime), and behave liquid-like:
low sound velocity and no transverse-like dynamics identifiable in
the longitudinal current. On the contrary the modes with frequency
satisfying $\Omega (Q) \tau_\alpha >> 1$ see a frozen, solid-like,
structure and behave as they were in the glassy phase: elastic
response, high sound velocity and identifiable transverse-like
dynamics.

A different interpretation for the origin of the two excitations
appearing in the $S(Q,\omega)$ of water has recently been proposed
by Petrillo et al.  \cite{8} and by Sacchetti et al.
\cite{sacchnew}, who analyzed the INS and IXS data using a
solid-like framework: the mode-mode interaction and the symmetry
avoided crossing. In this context, the second dispersion-less
feature is assigned to a local inter-molecular vibration,
resembling one of the optic modes in ice \cite{9e}. Moreover, at
variance to Ref.~\cite{25,10}, where the longitudinal sound
velocity dispersion is associated to the interaction of the
longitudinal sound waves with a relaxation process, the transition
from the hydrodynamic to the fast sound is attributed to the
interaction between the sound waves and this optic-like mode. As
the symmetry of the two modes is supposed to be the same, the two
branches repel each other to avoid crossing, and the high
frequency branch at large $Q$ disperses with a slope larger than
that of the low frequency branch at small $Q$. This model
describes the INS and IXS spectra as well as the viscoelastic
model; a definitive conclusion on the correct interpretation of
the water dynamics can therefore not be drawn on the basis of a
"best fit" procedure. Such a conclusion needs to rely on an
experimental basis, as that brought, for example, from the
physical meaning of the parameters entering in the viscoelastic
and in the solid-like models, and/or from their dependence on the
thermodynamic state.

In this paper we present new IXS data which significantly extend
the thermodynamic region investigated in previous IXS and INS
experiments. Moreover, thanks to the increased performance of the
IXS instrument, data with better statistical quality could be
recorded, allowing an even more reliable extraction of the
relevant parameters. The previous IXS study by Monaco et al.
\cite{10} was focused on the low Q-region (mainly below 7
nm$^{-1}$) where the second excitation does not give a significant
contribution, and a consistent viscoelastic analysis of the whole
spectrum could be performed. The present work covers a much larger
$Q$-range (from 4 to 30 nm$^{-1}$) and thermodynamic region by a
choice of pressure and temperature such that the relaxation time
$\tau_\alpha$ changes by more than a decade. At variance with the
previous works, this allows us to follow the evolution of the
second weakly dispersing feature not only as a function of
momentum transfer, but also all the way from the viscous to the
elastic limit. The extended thermodynamic and $Q$ regions and the
improved statistics allow us to provide overwhelming evidence in
favor of \textit{i)} a ''viscous'' - rather than solid-like -
origin of the transition from "normal" to "fast" sound in liquid
water and \textit{ii)} the transverse origin of the low frequency
weakly dispersing feature appearing in the $S(Q,\omega)$. The
paper is organized as follows: In section II we present the
experimental set-up and the information pertinent to the sample
cell, the theoretical formalism utilized in the data analysis, and
the experimental results. Section III provides the data analysis
and the discussion, while Section IV draws the conclusions.

\section{Experiment}
\subsection{Experimental setup}
The experiment was carried out at the Inelastic X-ray Scattering
Beamline II (ID28) at the European Synchrotron Radiation Facility
in Grenoble/France. The X-rays from an undulator source are
monochromatized by a cryogenically cooled silicon (111) double
crystal monochromator and a high-energy resolution backscattering
monochromator, operating at the silicon (11 11 11) reflection
order. The backscattered photons of energy 21.747 keV impinges on
a gold-coated toroidal mirror, which provides a focal spot at the
sample position of 270 (horizontal) and 80 (vertical) $\mu$m$^{2}$
Full Width at Half Maximum (FWHM). The incident flux on the sample
is 5$\times$10$^{9}$ photon/s. The scattered photons are
energy-analyzed by a Rowland circle five-crystal spectrometer,
operating at the same reflection order as the monochromator. The
energy-analyzed photons are detected by a Peltier-cooled silicon
diode detector which has an intrinsic energy resolution of 400 eV
\cite{14}. The dark counts due to electronic and environmental
noise amounts to about 0.003 counts/s. The momentum transfer $Q =
2k_{i} \cdot sin(\theta_{s}/2) $, where $k_i$ is the incident
photon wave vector and $\theta_s$ is the scattering angle, is
selected by rotating the spectrometer around a vertical axis
passing through the scattering sample in the horizontal plane.
Since there are five independent analyzer systems, spectra at five
different momentum transfers can be recorded simultaneously. Their
separation amounts to approximately  3 nm$^{-1}$ for the Si (11 11
11) reflection. The energy scans are performed by varying the
monochromator temperature while the analyzer temperature is kept
fixed. Conversion from the temperature scale to the energy scale
is accomplished by the following relation: $\Delta E/E = \alpha
\Delta T$, where $\alpha = 2.58 \cdot 10^{-6}$ is the linear
thermal expansion coefficient of silicon at room temperature. The
validity of this conversion has been checked by comparing the
measured diamond dispersion curve for longitudinal acoustic and
optical phonons with well established inelastic neutron scattering
results. The overall experimental resolution is experimentally
determined by measuring the scattering from a disordered sample of
plexiglass at a Q-transfer of 10 nm$^{-1}$, corresponding to the
first maximum in the static structure factor $S(Q)$, and at T = 10
K in order to maximize the elastic contribution to the scattering.

Distilled and deionized water was loaded into a specially designed
stainless steel cell, made out of INCONEL-751, which can be
pressurized by a hand pump up to 5 kbar and heated up to 700 K.
The X-ray beam passes through two 1-mm-thick diamond single
crystal windows with a 2.3 mm aperture. The distance between the
two windows, i.e., the sample length along the X-ray beam was 10
mm. The cell geometry allowed us to cover the relevant momentum
transfer regime up to 30 $ nm^{-1} $ with a sample length
comparable to the X-ray photoabsorption length ($\mu \approx$0.1
mm$^{-1}$ at $E$=21 keV). The high pressure cell was kept in
vacuum in order to minimize both temperature gradients and air
scattering. The pressure was monitored with a calibrated gauge
whose precision is about 1$\%$. The temperature was determined by
a Cr-Al thermocouple in direct contact with the main body of the
cell with an accuracy of $\pm$ 1 K. Table \ref{points} reports the
thermodynamic conditions of the experiment as well as the
respective shear viscosity ($\eta$) and the adiabatic speed of
sound ($c_0$) values.
\begin{table}[htbp]
\begin{ruledtabular}
\begin{tabular}{ccccc}
  T & P & $\rho$ & $c_0$ & $\eta$ \\
  (K) & (bar) & ($g/cm^3$) & (m/s) & (cP) \\\hline
  263 & 2000 & 1.084 & 1825 & 2.09 \\
  282 & 2000 & 1.078 & 1820 & 1.32 \\
  296 & 2000 & 1.073 & 1840 & 0.96 \\
  321 & 2000 & 1.072 & 1875 & 0.62 \\
  359 & 2000 & 1.041 & 1880 & 0.38 \\
  419 & 95   & 0.926 & 1500 & 0.19 \\
\end{tabular}
\end{ruledtabular}
\caption{Thermodynamic conditions of temperature T and pressure P
at which the IXS spectra were recorded. The uncertainty onf $T$ is
$\pm$1 K and that of $P$ is $\pm$1 \%. Columns 3 -5 report the
three corresponding thermodynamic quantities used in the data
analysis, namely the mass density $\rho$, the adiabatic sound
velocity $c_0$, and the shear viscosity $\eta$. Their values were
obtained from the water Equation of State \cite{16}. The freezing
point of water at $P$=2 kbar is 251 K.}\label{points}
\end{table}
\subsection{Theoretical formalism and fitting procedure}
The theoretical framework used to interpret the experimental
spectra is based on the generalized Langevin equation (memory
function approach), more specifically on the generalized
hydrodynamic description. In this framework the formal structure
of the "classical" hydrodynamic regime is retained, but the
thermodynamic derivatives and transport coefficients are replaced
by functions which can vary both in space (or wave number) and
time (or frequency) \cite{17,18}. In the last years, this approach
has been successfully applied to the interpretation of the high
frequency dynamics of different systems, including water, and
spanning from noble gases \cite{n1} to liquid metals \cite{n2}, as
well as from supercooled liquids \cite{n3} to glasses
\cite{sio2b}. The basic equations are discussed in detail in
Ref.~\cite{10}; here, we only stress the key points and the
details relevant to the present data analysis.

The longitudinal dynamics is described by a Langevin equation for
the density-density correlation function $F(Q,t)$:
\begin{equation}
\frac{\partial^2 F(Q,t)}{\partial t^2}+\omega_0^2(Q) F(Q,t) +
\int_0^t dt' m_Q(t-t') \frac{\partial F(Q,t')}{\partial t'}=0;
\label{Langevin}
\end{equation}
and, consequently, the expression for the $S(Q,\omega)$
(normalized to $S(Q)$), the time Fourier transform of $F(Q,t)$,
reads:
\begin{equation}
\frac{S(Q,\omega)}{S(Q)}=\frac{1}{\pi}
\frac{\omega_{0}^2\tilde{m}_Q'(\omega)}{(\omega_{0}^2-\omega^2
-\omega\tilde{m}_Q''(\omega))^2+(\omega\tilde{m}_Q'(\omega))^2}
\label{Sqw}
\end{equation}
where $\tilde{m}'_Q(\omega)$ and $\tilde{m}''_Q(\omega)$ are the
real and imaginary part of the Fourier transform of the memory
function $m_Q(t)$. Its generalized form can be written as:
\begin{equation}
m_Q(t)=\omega_o^2(Q) [\gamma(Q)-1] e^{-D_T(Q) Q^2 t} + K_L(Q,t),
\label{memGen}
\end{equation}
where $\omega_o^2=K_{B}T Q^2/(M S(Q))$ is the (squared) frequency
of the sound waves excitation in the hydrodynamic regime, $\gamma
(Q)$ and $D_T(Q)$ are the Q-dependent generalizations of $\gamma =
C_p/C_v$ (the constant pressure to constant volume specific heat
ratio) and $D_T=\kappa /(\rho C_v)$, $\kappa$ is the thermal
conductivity, $S(Q)$ is the static structure factor, $M$ is the
molecule mass and $K_{B}$ the Boltzmann constant. The second
contribution to the memory function, $K_L(Q,t)$, a quantity
directly related to the longitudinal kinematic viscosity
\cite{17,18}, is here chosen as:
\begin{equation} K_L(Q,t)=2
\Gamma_\mu(Q) \delta(t) + \Delta^2(Q) e^{-t/\tau_\alpha (Q)},
\label{modello}
\end{equation}
where 2$\Gamma_\mu (Q)\delta(t)$ represents a fast (microscopic or
$\mu$) decaying contribution to $K_L(Q,t)$ and the exponential
term accounts for the structural ($\alpha$) relaxation process.
$\tau_\alpha$ is the $Q$-dependent time which characterizes the
long time tail of $K_L(Q,t)$, and $\Delta^{2}(Q)$ is the
structural relaxation strength which is related to $c_\infty$(Q)
and $c_0$(Q) - the $Q$-dependent generalizations of the usual
infinite frequency and adiabatic sound speeds \cite{20} via
$\Delta^{2}=(c_\infty-c_0)Q^2$. In summary, the resulting
expression for the memory function becomes:
\begin{equation} \nonumber
m_Q(t)=\omega_o^2(Q) (\gamma(Q)-1) e^{-D_T(Q) Q^2 t} + 2
\Gamma_\mu(Q) \delta(t) + \Delta^2(Q) e^{-t/\tau_\alpha(Q))}.
\label{memmod}
\end{equation}
It is worth pointing out that the time dependence of the
structural relaxation process contribution to the memory function
is assumed here to be Debye-like (no stretching is present). This
is in apparent contrast to recent findings \cite{torre,masciov}
where the existence of a non-negligible stretching ($\beta \sim
0.6$) in the correlation function of supercooled water was
observed. However, as also pointed out in \cite{masciov}, the
existence of stretching is hardly detectable in the temperature
region investigated here. As a matter of fact, we did not find any
significant statistical improvement in introducing a new fitting
parameter ($\beta$). We therefore used a simplified version of the
memory function and fixed $\beta$ to one.

\begin{figure}[t]
\includegraphics[width=.5\textwidth]{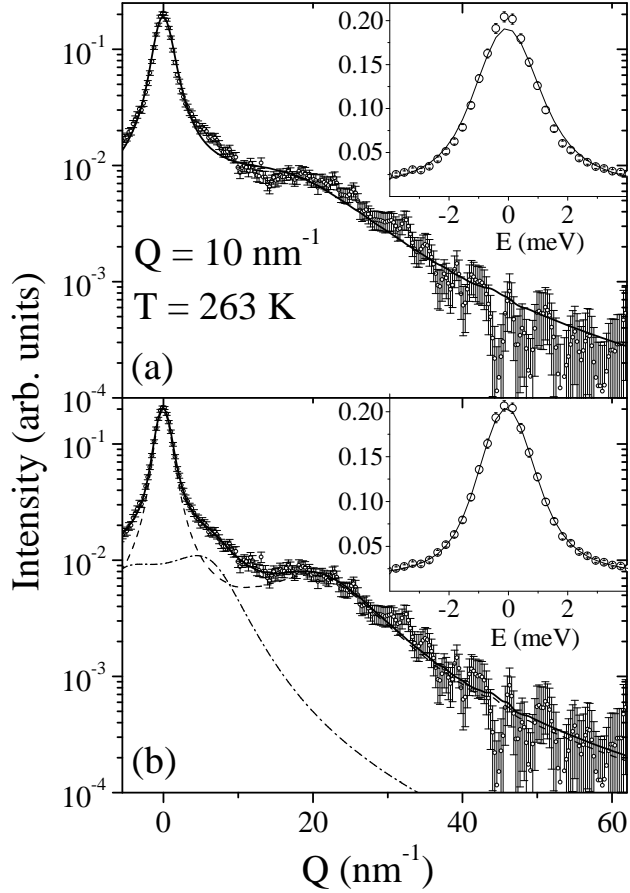}
\caption{\footnotesize{IXS spectrum at $Q$=10 nm$^{-1} $, T=263 K
and P=2 kbar. The experimental data and their error bars are shown
together with the best fits utilizing the model function described
in the text with and without the second inelastic contribution.
Panel (a): one viscoelastic contribution. Panel (b): one
viscoelastic contribution and one damped harmonic oscillator
model. The inset shows the central peak region.}} \label{cfr}
\end{figure}

The viscoelastic analysis of the water IXS spectra previously
performed \cite{10} was confined to $Q$-values not exceeding $Q$=7
nm$^{-1}$, a region in which the second excitation discussed in
the introduction is not visible. Consequently, the applied
theoretical formalism yielded excellent agreement between the
model function and the experimental data. In the present case, the
inclusion of a second excitation is mandatory as it is evident,
for example, by inspection of Figure \ref{cfr}, which reports the
IXS spectrum recorded at $T$=263 K, $P$=2 kbar and $Q$=10
nm$^{-1}$. In this figure, panel (a) shows the best fit to the
data with only one excitation (viscoelastic model) according to
Eq.~\ref{Sqw}. Both the inelastic and the central part of the
spectra are not well reproduced, and it is obvious that a second
excitation has to be added. This has been done for the fit
presented in panel (b) where the second excitation is accounted
for by a damped-harmonic oscillator (DHO) function \cite{21} {\it
added} to the viscoelastic $S(Q,\omega)$, resulting in a
significantly better fit. It is important to stress that this is a
purely empirical approach to introduce the mixing phenomenon of
longitudinal and transverse dynamics. A correct theoretical
description should provide a memory function, directly reproducing
a double-excitation line shape, which naturally would involve a
much larger set of variables. In particular, such a formalism
should be extended to a regime of broken ergodicity, when the
frequency range is much higher than the structural relaxation time
($\omega \tau_\alpha > 1$). In this range the symmetry arguments
that decouple the transverse (T) from the longitudinal (L)
variables in the generalized Langevin equation must be abandoned
and a coupling $L-T$ mechanism must be taken into account. Work
along these lines is currently in progress. Within the present
context, the choice of a DHO lineshape to describe the transverse
dynamics is used empirically to provide information on the
intensity and the frequency of the second excitation. In summary,
the model function used to represent the spectra is composed of
the following pieces:
\begin{enumerate}
\item[(1)] A viscoelastic model function, proportional to
Eq.~\ref{Sqw}, to account for the central peak and the
longitudinal dynamics: $$ s^{(L)}(Q,\omega)\;=\;
\frac{A_L}{\pi}\frac{\omega_{0}^2\tilde{m}_Q'(\omega)}
{(\omega_{0}^2-\omega^2
-\omega\tilde{m}_Q''(\omega))^2+(\omega\tilde{m}_Q'(\omega))^2}$$
We have neglected the contribution due to thermal relaxation,
which amounts to set $\gamma (Q) = 1$, an approximation which
turns out to be very good in the high Q region. \item[(2)] A DHO
line-shape for the second excitation:
$$ s^{(T)}(Q,\omega)\;=\;\frac{A_T}{\pi}
\frac{\Omega_{T}^2\Gamma_T}{(\Omega^{2}_T-\omega^2)^2
+(\omega\Gamma_T)^2}$$ Where $\Omega_T$ is the maximum of the
"transverse-like" contribution to the the longitudinal current
spectrum (the dynamic structure factor multiplied by
$\omega^2/Q^2$, i.~e. of $\omega^2 s^{(T)}(Q,\omega)$) and
$\Gamma_T$ is the width of the inelastic transverse peaks.
\end{enumerate}
This model function for the dynamic structure factor is symmetric,
and, in order to account for the quantized character of the energy
transfers at the microscopic level, has to be weighted with a
function that i) satisfies the detailed balance and ii) becomes
unity in the classical limit ($\hbar\omega/K_BT<<1$). Among the
possible choices, the weighting factor usually utilized is:
$$
w(\omega,T)=\frac{\hbar\omega}{K_B T}(n(\omega,T)+1)=
\frac{\hbar\omega}{K_B T}[1-exp(-\hbar\omega / k_B T)]^{-1}
$$
Finally, the theoretical model has to be convoluted with the
experimental resolution function $R(\omega)$, to give a fitting
function of the form:
\begin{eqnarray}
I(Q,\omega)\;=\;\{R(\omega)\}\; \otimes \;
\{\frac{\hbar\omega}{K_BT}[n(\omega,T)+1]
[s^{(L)}(Q,\omega)+s^{(T)}(Q,\omega)]\}\;+\;B
\label{Integ}
\end{eqnarray}
Here B is an additional term which accounts for the electronic and
environmental  background of the detectors. As a result, the data
are fitted with 9 free parameters: i) the background $B$; the
transverse ii) "intensity" $A_T$, iii) position $\Omega_T$ and iv)
width $\Gamma_T$; v)the longitudinal intensity $A_L$; the three
parameters of the longitudinal memory function, namely the
structural relaxation vi) time $\tau_\alpha$ and vii) strength
$\Delta$ and the area of the microscopic relaxation $\Gamma_\mu$;
and, finally, ix) the static structure factor $S(Q)$ entering in
the term $\omega_o$. All these parameters are, in principle, $T$-
and $Q$-dependent.

As a matter of fact, $\Gamma_\mu$, which represents the
topological microscopic disorder contribution to the acoustic
attenuation, is known to have a negligible temperature dependence.
Therefore we have fixed its value to the one obtained at the
lowest temperature. We checked that leaving this parameter free
does not change significatively the values of the other
parameters. Moreover, we observe that, for the highest temperature
spectra, leaving the energy position and width of the transverse
contribution as free parameters in the fit, results in a
meaningless overdamping of the DHO function to the extent that it
cannot be distinguished from the purely relaxational elastic
contribution already accounted for by the viscoelastic model (the
intensity of this contribution vanishes at high $T$, thus implying
that $\Omega_T$ and $\Gamma_T$ become irrelevant). In order to
circumvent this, and guided by the observation of Sokolov et al.
\cite{sok}, the position and width of the second excitation was
unambiguously determined in an unconstrained fit for the lowest
temperature IXS spectra, and kept fixed at higher temperatures.
Therefore, at all temperature but the lowest, there are six free
fitting parameters.

\subsection{Experimental results}

\begin{figure*}[t]
\hspace{-5cm}
\includegraphics[width=.75\textwidth]{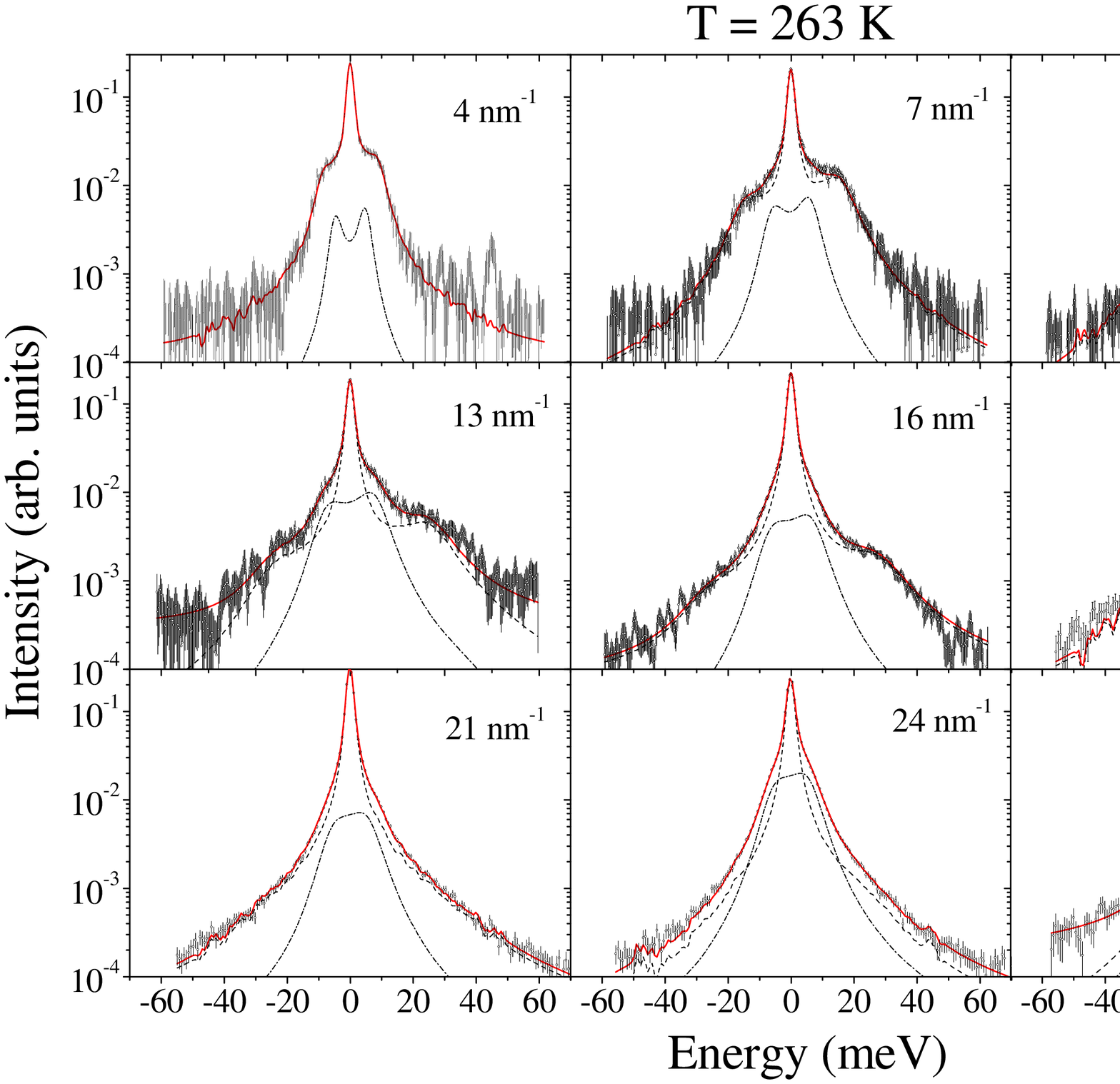}
\caption{\footnotesize{IXS spectra of water at $T$=263 K and $P$=2
kbar. The figure reports the data (dots) normalized to the total
integrated intensity on a logarithmic scale emphasizing the
inelastic part of the spectrum. The solid line represents the
total fit result as explained in the text, while the other lines
visualize the two contributions to the fitting model: the
longitudinal and quasi-elastic one represented by a viscoelastic
model (dashed line) and the transverse-like secondary peak
accounted for by a DHO model function (dash-dotted line)}}
\label{263log}
\end{figure*}

\begin{figure*}[t]
\hspace{-5cm}
\includegraphics[width=.75\textwidth]{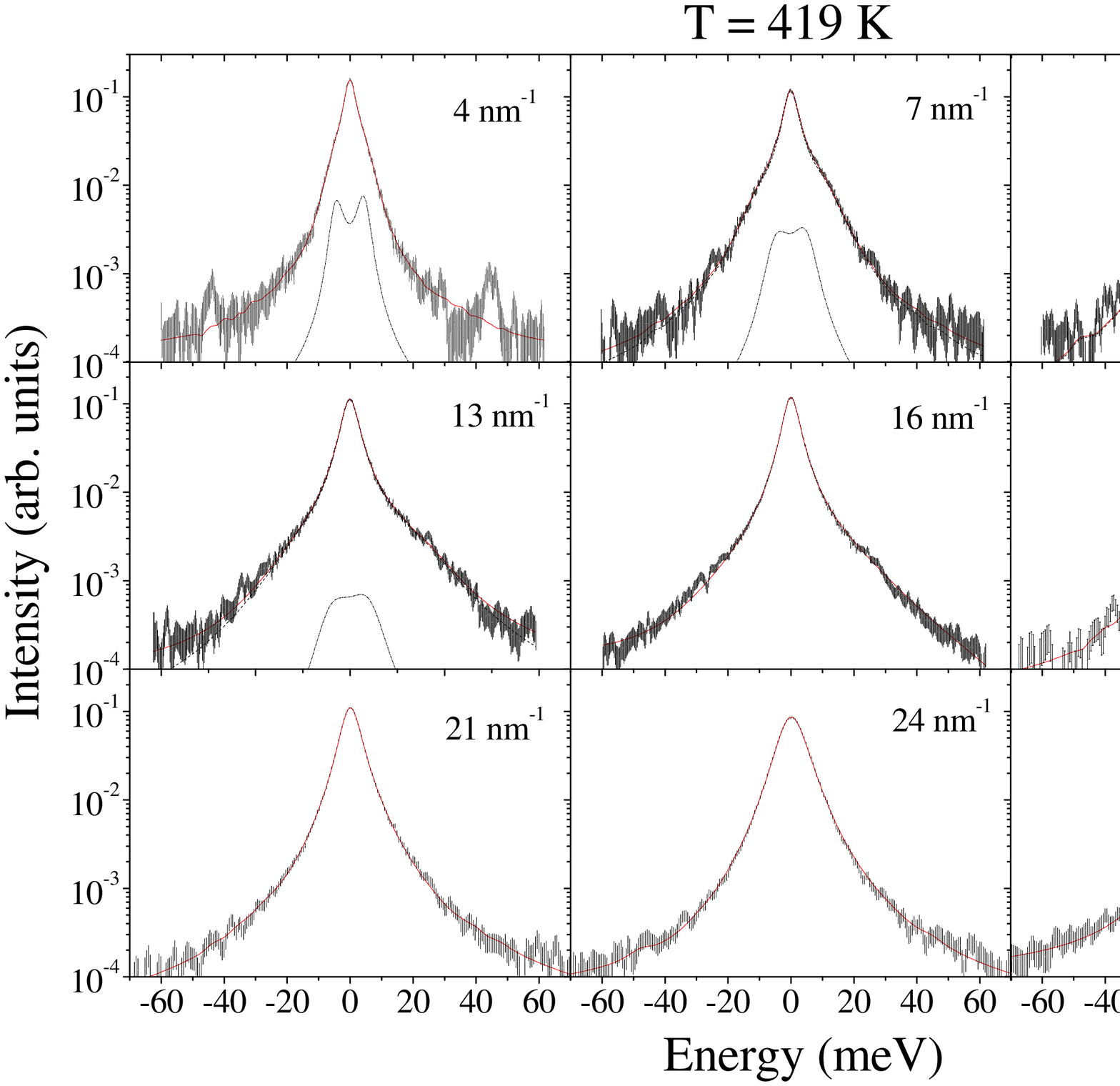}
\caption{\footnotesize{IXS spectra at $T$=419 K and $P$=95 bar.
The figure reports the data (dots) normalized to the total
integrated intensity on a logarithmic scale emphasizing the
inelastic part of the spectrum. The solid line represents the fit
result as explained in the text, while the other lines visualize
the two contributions to the fitting model: the longitudinal and
quasi-elastic one represented by a viscoelastic model (dashed
line) and the transverse-like secondary peak accounted for by a
DHO model function (dash-dotted line)}} \label{419log}
\end{figure*}

\begin{figure*}[htbp]
\hspace{-5cm}
\includegraphics[width=.75\textwidth]{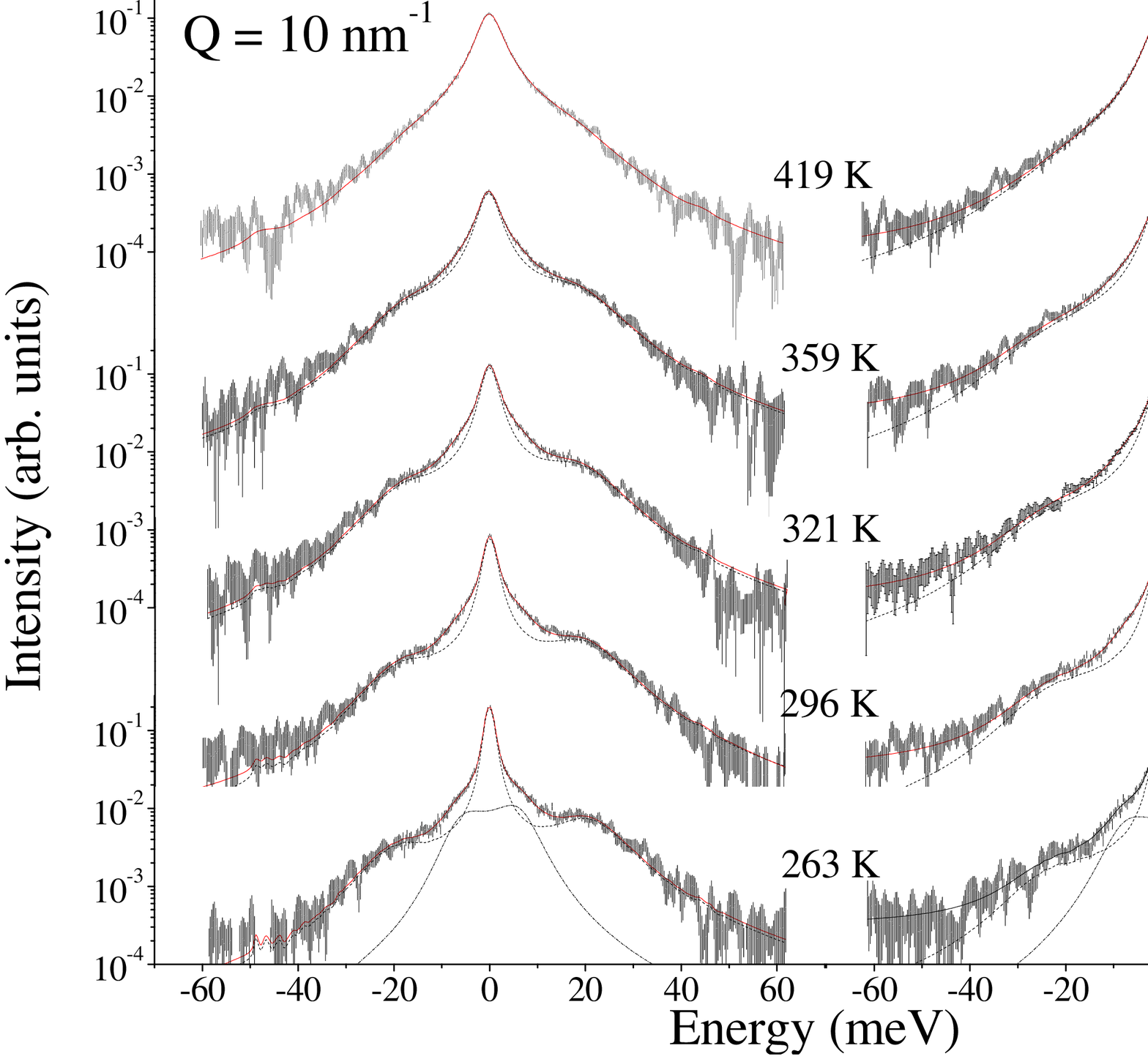}
\caption{\footnotesize{Temperature evolution of the IXS spectra
for $Q$=10 $nm^{-1} $ (left panel) and $Q$ = 13 $nm^{-1} $ (right
panel). The figure reports the data (dots) normalized to the total
integrated intensity on a logarithmic scale emphasizing the
inelastic part of the spectrum. The solid line represents the fit
result as explained in the text, while the other lines visualize
the two contributions to the fitting model: the longitudinal and
quasi-elastic one represented by a viscoelastic model (dashed
line) and the transverse-like secondary peak accounted for by a
DHO model function (dash-dotted line).}}\label{Tlog}
\end{figure*}
IXS spectra were recorded at the different temperatures and
pressures as indicated in Table \ref{points}, spanning the
momentum transfer, Q, region from 4 to 16 nm$^{-1}$ with an
approximate constant spacing of 3 nm$^{-1}$. For the lowest (263
K) and highest (419 K) temperatures, the  Q-range was extended to
30 nm$^{-1}$. The spectra extend up to 60 meV on both sides of the
elastic line, with an integration time of 100 s per spectrum. For
each setting three spectra were collected, and subsequently summed
in order to achieve the high statistical accuracy required for the
data analysis. Typical total counts range between 1500 (at 4
nm$^{-1}$), 2500 (at 9.93 nm$^{-1}$), and 3500 (at 12.91
nm$^{-1}$). To account for the slow drift of the photon flux
impinging  on the sample, the collected data were normalized to
the intensity of the incident beam. A typical IXS spectrum is
shown on a logarithmic scale in figure \ref{cfr}. The experimental
data with their error bars are shown together with the best fits
(full lines) using both the model function discussed before
(bottom panel) and the same function with $A_T$=0 (top panel). For
clarity, only the elastic and the Stokes part of the spectrum is
shown. The insets provide a zoom of the elastic line region on a
linear scale. It can be easily appreciated that a fit using a
viscoelastic model only (top panel), does not properly describe
the IXS data, leading to a poor fit of  both the elastic line and
the inelastic part of the spectrum (the reduced $\chi^2$ results
to be $\approx$ 2, i.~e. more than 10$\sigma_{\chi^2}$ larger than
its expected value). The inclusion of a second inelastic
excitation significantly improves the quality of the fit (reduced
$\chi^2 \approx 1$, $\sigma_{\chi^2}\approx 0.1$), as can be seen
in the lower panel. For all fitted spectra, the value of $\chi^2$
remained within one standard deviation from the expected value.

Figure \ref{263log} shows as an example the $Q$ evolution of the
IXS data recorded at T=263 K and P=2000 bar. The two inelastic
contributions which clearly appear at $Q$=10 nm$^{-1}$ can also be
seen throughout the whole Q-range. As can be noticed by the $Q$
evolution of the dashed and dotted lines (longitudinal and
transverse-like contributions, respectively) the intensity of the
low-frequency (transverse-like) mode increases with increasing
$Q$.

The shape of the IXS spectra as well as their $Q$-dependence is
markedly different at $T$=419 K and $P$=95 bar (see figure
\ref{419log}). The elastic line appears broader, and consequently
the low frequency, non dispersing mode is much less visible. As a
matter of fact, it can only be identified in the spectra with $Q$
$\leq$ 13 nm$^{-1}$, while for larger Q values the increasing
width of the elastic line completely governs the spectral shape.
The weak feature at about 45 meV in the spectra at $Q$=4 nm$^{-1}$
is the contribution from the high-pressure cell windows and
corresponds to the diamond longitudinal acoustic phonon.

To show an example of the temperature evolution of the IXS
spectra, and more specifically, of the behavior of the second
weakly dispersing feature, the spectra recorded at $Q$=10 (left
panel) and 13 nm$^{-1}$ (right panel) are shown in figure
\ref{Tlog} at the indicated temperatures. At the lowest
temperature, the experimental data are shown together with the
best fit to the model function and the two individual components
(viscoelastic model and DHO). For the higher temperatures, only
the viscoelastic contribution and the total fit are shown for the
sake of clarity and in order to emphasize the decreasing
importance of the second excitation. We notice a clear trend for
both momentum transfers: with increasing temperature the
contribution of the second excitation becomes smaller and smaller
to the extent that it completely disappears at the highest
temperature.

\section{Data analysis and discussion}
\subsection{The longitudinal dynamics}

\begin{figure}[t]
\vspace{0cm}
\includegraphics[width=.75\textwidth]{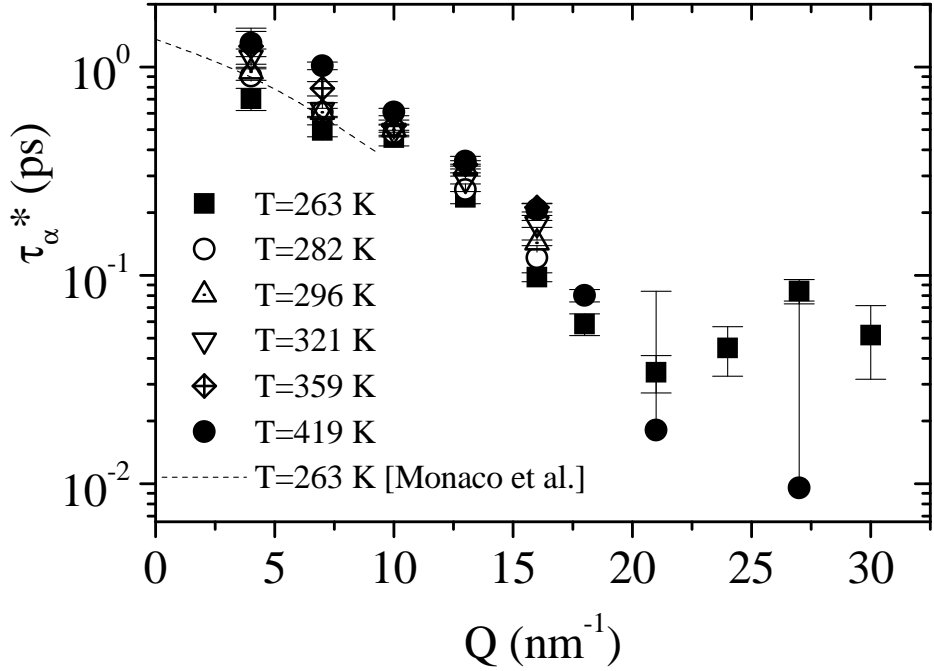}
\vspace{0cm} \caption{\footnotesize{$Q$-dependence of the (scaled)
structural relaxation time $ \tau_{\alpha}^{\star} = \tau_\alpha
\times \eta(T=263 K) / \eta(T)$ (see text). The dashed line
reports the fit results of \cite{10} for the Q dependence of the
relaxation time at $T=263 K$, valid up to Q = 7 $ nm^{-1}$.}}
\label{tvsQ}
\end{figure}

\begin{figure}[t]
\includegraphics[width=.75\textwidth]{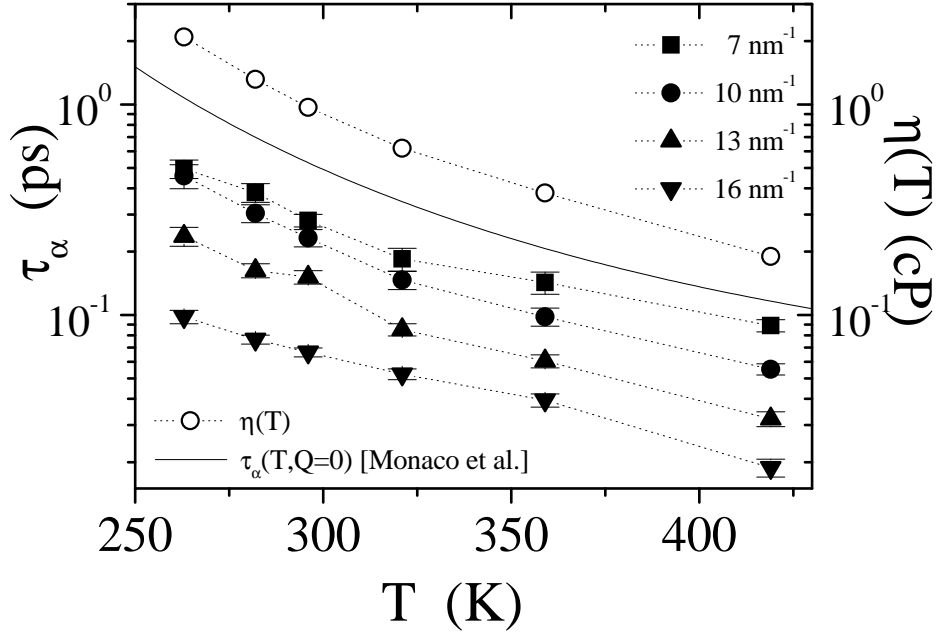}
\vspace{0cm} \caption{\footnotesize{Temperature evolution of the
structural relaxation time (full symbols) at different $Q$-values
between 7 and 16 $ nm^{-1} $. The shear viscosity $\eta(T)$ is
reported for comparison (open dots), as well as the $Q=0$
extrapolation of the relaxation time \cite{10} (solid line).}}
\label{tvsT}
\end{figure}

In the present section we present and discuss the results obtained
for the longitudinal dynamics which are retrieved from the
viscoelastic analysis as described in detail in section II B.

In figure \ref{tvsQ} we report the $Q$-dependence of the
structural relaxation time $\tau_\alpha$ at different temperatures
$T$, scaled by the shear viscosity ratio $r(T)=\eta(T=263 K) /
\eta(T) $: i.~e. $\tau_{\alpha}^{\star}$ = $r(T)
\tau_{\alpha}(T)$. This scaling has been performed in order to
verify whether the structural relaxation time is proportional to
the longitudinal viscosity $\eta$, a relationship which generally
holds in the continuum limit $Q \longrightarrow 0 $. It is evident
from the inspection of figure \ref{tvsQ} that the $Q$ evolution of
$\tau_{\alpha}$ at different T is indeed the same, especially in
the $Q$-range below 20 nm$^{-1}$. As implied by the validity of
the scaling, the $T$ and the $Q$ dependence of the relaxation time
can be factorized: $\tau_\alpha(Q,T) = \eta(T) f(Q)$, being $f(Q)$
the function that describes the $Q$-evolution of the relaxation
time. The dashed line at low $Q$ in Fig. \ref{tvsQ} reports the
dependence of the relaxation time for $T$=263 K, derived in
Ref.~\cite{10}; it turns out to be consistent with the present
results.  Figure \ref{tvsT} shows the temperature dependence of
$\tau_\alpha$ for different $Q$-values between 7 and 16 nm$^{-1}$.
The evolution of the viscosity (open circles) and the $Q
\rightarrow 0$ limit of $\tau_\alpha(Q,T)$ (solid line) as
reported in Ref.~\cite{10} are also shown. The good overall
agreement confirms the reliability of the structural relaxation
time determination by the viscoelastic model over the large $Q$
range and in the different thermodynamic conditions investigated
here.

\begin{figure*}[t]
\hspace{-5cm}
\includegraphics[width=.75\textwidth]{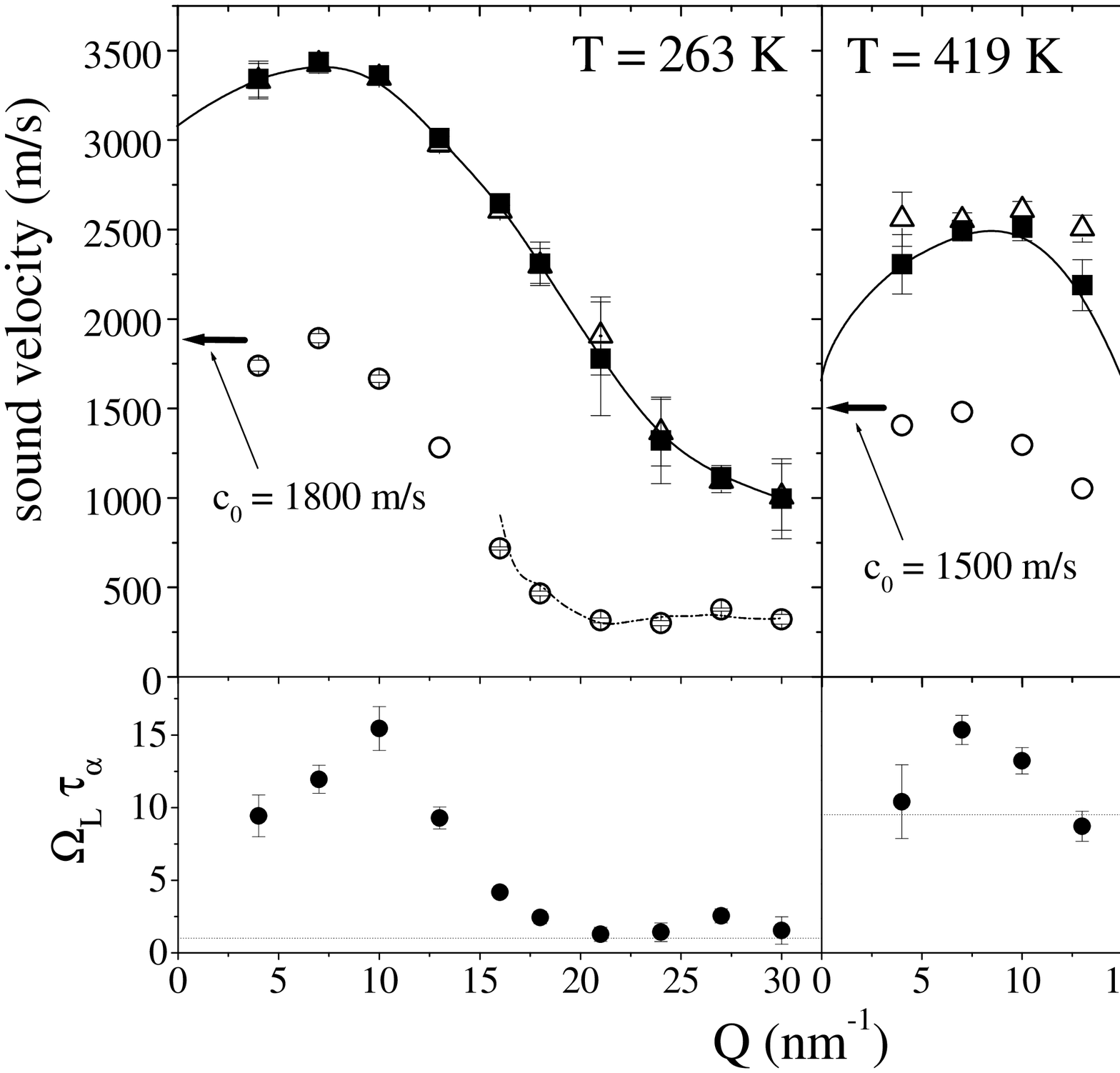}
\caption{\footnotesize{Sound velocities and the product $\Omega_L
\tau_\alpha$ -as determined by the fit to the IXS spectra- as a
function of $Q$, and at two different temperatures: $T$=263 K
(left panels) and $T$=419 K (right panels). The top panels show
the $Q$-evolution of the sound velocities $ c_{0} $ (open dots), $
c_{\infty} $ (open triangles), and $ c_{L} $ (full squares),
whereas the bottom panels show the product $\Omega_L \tau_\alpha$,
that, when equal to unity, marks the transition from the infinite
frequency sound velocity $c_\infty$ to the zero-frequency sound
velocity $c_0$. The solid line is a guide to the eye. The
hydrodynamic ($Q$=0) value of $ c_{0} $ is indicated by arrows;
the Q-evolution of $ c_{0}(Q) $, as derived using Eq.~\ref{ccc}
and the $S(Q)$ determined by neutron scattering, is indicated by
the dash-dotted line in the $Q$ region where neutron data are
available. The dotted horizontal line in the lower panels marks
the condition $\Omega_L \tau_\alpha$ = 1.}} \label{svQ}
\end{figure*}

The temperature dependence of the zero frequency sound velocity
($c_0 = \omega_0 /Q$), the infinite frequency sound velocity
($c_\infty =\sqrt{(\Delta^2 + \omega_0^2)} /Q$) and the apparent
sound velocity ($c_L=\Omega_L/Q$) are reported in figure \ref{svQ}
for $T$=263 K (left panel) and $T$=419 K (right panel). Here,
$\Omega_L$ is the maximum of the function $\omega^2
s^{(L)}(Q,\omega)$ calculated using the best fit parameters. The
bottom panels of Fig.~\ref{svQ} show the $Q$-dependence of the
product $\Omega_L \tau_\alpha$, the parameter that indicates
whether the dynamics is viscous-like ($\Omega_L \tau_\alpha<1$) or
elastic-like ($\Omega_L \tau_\alpha>1$). For $T$=263 K, $\Omega_L
 \tau_\alpha$ is always larger than, or close to, one, the system has a
solid-like response, and the sound velocity $c_L$ is close to
$c_\infty$, its infinite frequency value, over the whole explored
$Q$-range. In contrast, at $T$=419 K, $\Omega_L \tau_\alpha$ is
larger than one only over a limited $Q$-range, from 4 to 10
nm$^{-1}$, and then decreases rapidly below one for increasing
$Q$-values. Consequently, we observe a transition from the
infinite- to the zero-frequency sound regime: it takes place for
$Q$ values between 10 and 15 nm$^{-1}$. This is the first
experimental observation of the transition $c_\infty \rightarrow
c_0$ that takes place for $Q$-values around the $Q$'s where the
$S(Q)$ shows its first maximum, as a consequence of the De~Gennes
narrowing.The $Q$ dependence of the excitation frequency, in fact,
shows a decrease with a minimum just in correspondence with the
first maximum of the $S(Q)$ as a memory of a "pseudo-Second
Brillouin zone". At the same time the relaxation time decreases
towards high $Q$ with respect to the $Q=0$ value with a slight
increase at around 27 nm$^{-1}$ again due to De~Gennes narrowing
(see figure \ref{tvsQ}). Such kind of $c_\infty \rightarrow c_0$
transition, taking place at $Q$ values around the first maximum of
the $S(Q)$, has been already observed in MD simulations of a
Lennard-Jones model glass \cite{nonh}. Furthermore, we note that
our derived values for $c_0$ is in excellent agreement with
independent determinations, both in the limit of $ Q \rightarrow 0
$ and at high $Q$-values. The values of $c_0$ in the low $Q$-limit
were obtained from thermodynamic data \cite{16}, and are indicated
by the arrows in Fig.~\ref{svQ}. The zero frequency sound velocity
in the high-$Q$ region (dash-dotted line) has been calculated
utilizing the expression for $c_0(Q)$ within the framework of
generalized hydrodynamics:

\begin{equation}
\label{ccc} c_0=\sqrt{\frac{K_{B}T}{M S(Q)}}
\end{equation}

Here, we utilized the $S(Q)$ values determined by neutron
diffraction measurements \cite{ricci} at the same thermodynamical
conditions (the $S_{OO}$ partial structure factor has been used).
The agreement of $c_0(Q)$, calculated using the previous relation,
and $c_0 (Q)$ derived from the fits convincingly demonstrates the
solidity of our data analysis procedure.

\begin{figure}[hbtp]
\includegraphics[width=.55\textwidth]{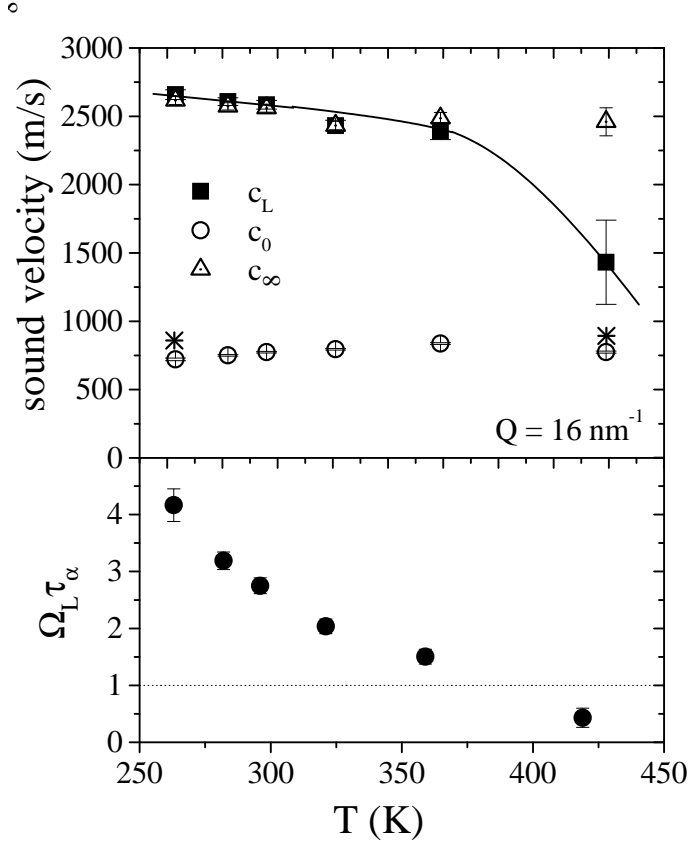}
\caption{\footnotesize{The temperature dependence of the sound
velocities and of the product $\Omega_L \tau_\alpha$ for $Q$=16
nm$^{-1}$. The data show the same phenomenology as those reported
in fig.~\ref{svQ}. Here the product $\Omega_L \tau_\alpha$ is
changed by varying the relaxation time scale: as in the previous
case we start to observe the transition when we approach $\Omega_L
\tau_\alpha \approx 1$. The solid line is a guide to the eye. The
stars in the upper panel indicate the $c_{0}(Q)$ values as derived
from Eq.~\ref{ccc} for the thermodynamical points available from
the neutron diffraction database \cite{ricci}.}}\label{svT}
\end{figure}

As expected for a viscoelastic behavior of the excitation, the
same $c_0 \leftrightarrows c_\infty$ transition can be observed at
fixed $Q$ due to the effect of the temperature on $\tau_\alpha$.
Figure \ref{svT} shows $c_L$, $c_0$ and $c_\infty$ as a function
of temperature for $Q$=16 nm$^{-1}$. At low $T$ the relaxation
time is long, $\Omega_L \tau_\alpha >> 1$, and the apparent sound
velocity is close to $c_\infty$. When, on increasing the
temperature, the structural relaxation enters in the excitation
timescale ($\Omega_L \tau_\alpha \approx 1$), the sound velocity
$c_L$ recovers its liquid-like value $c_0$.

\subsection{The second excitation}

\begin{figure}[htbp]
\includegraphics[width=.55\textwidth]{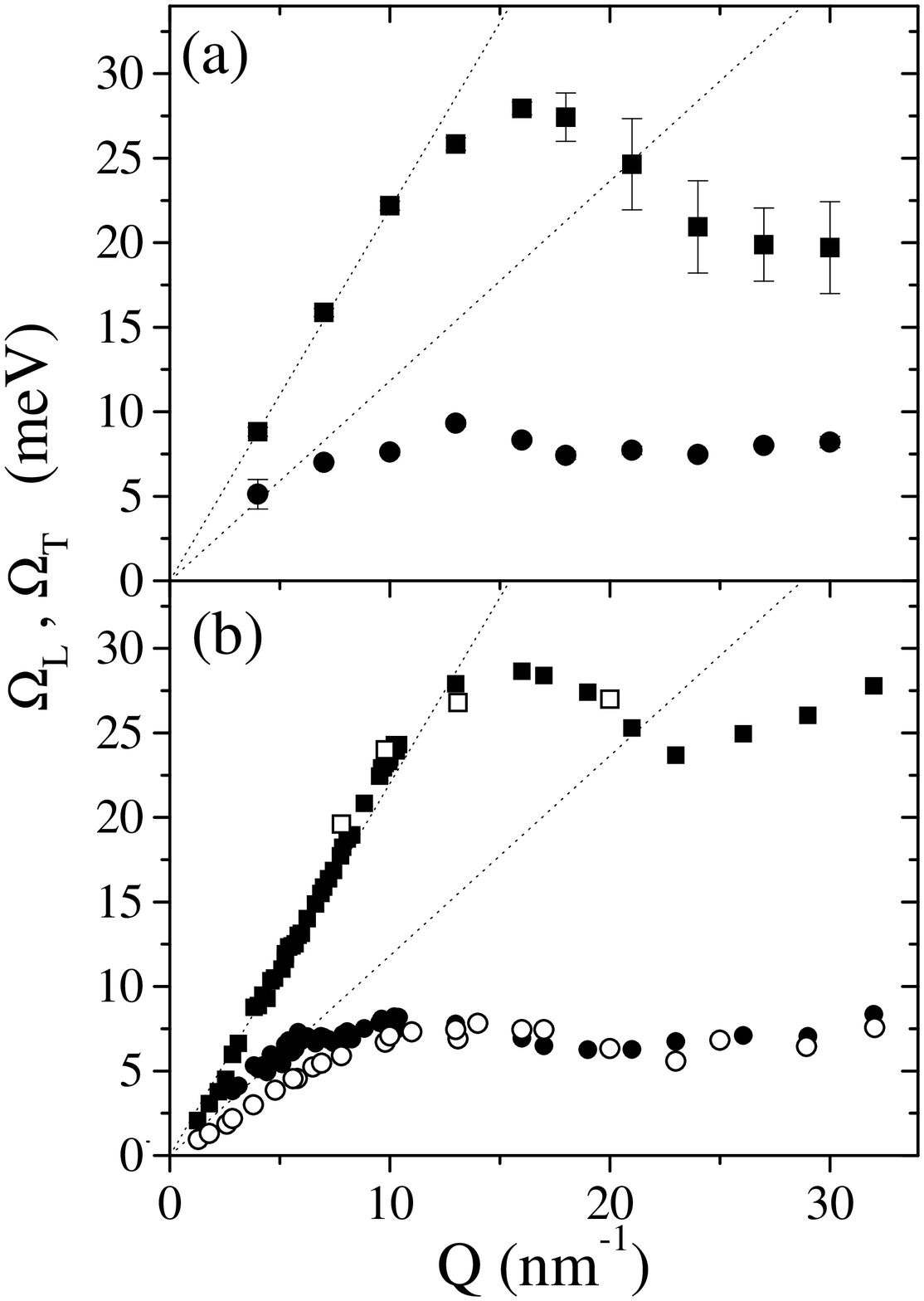}
\caption{\footnotesize{Panel (a): Longitudinal current peaks
positions from the fits of IXS data of liquid water at T=263 K and
P=2 kbar. Panel (b): MD simulations results \cite{22} (see text).
The dotted lines represent excitations with a linear dispersion in
Q, corresponding to a sound speed of 3300 and 1900 m/s,
respectively.}} \label{dispLT}
\end{figure}

As already discussed in some detail above, the IXS spectra,
especially at low temperature, can only be properly described if
one takes into account a second inelastic excitation. This second
excitation is described by three parameters (intensity, peak
position and peak width). Figure \ref{dispLT} shows the dispersion
($Q$-dependence) of $\Omega_T$, together with that of $\Omega_L$,
at T=263 K and P=2000 bar, where the second excitation is best
appreciated. The upper panel shows the result of the present work.
The longitudinal frequency $\Omega_L$ (squares) and the
"transverse" frequency $\Omega_T$ (circles) (both derived from the
fit procedure) show the expected behavior: a linear dispersion
followed by a bend down and a minimum in the $Q$ region where
$S(Q)$ has its maximum for the longitudinal excitation, and a weak
$Q$ dependence for the transverse mode. Full symbols in the lower
panel correspond to the results of a previous MD simulation
\cite{22} for the position of the Longitudinal current peaks.
These were carried out considering 4000 $ D_{2}O $ molecules
enclosed in a cubic box with periodic boundary conditions, and
utilizing the SPC/E model \cite{23}. The simulations were
performed at a density of 1 g/cm$^{3}$ and T $\approx$ 250 K. The
qualitative agreement between experiment and simulation is
remarkable. We note slight differences in the absolute values of
the excitation energies: while the maximum of the longitudinal
dispersion is slightly lower, the energies for the second
excitation are slightly higher in the case of the IXS results.
Moreover a small displacement at higher $Q$ in the position of the
first minimum in the IXS data is justified as these are taken at
high pressure while the simulations are performed at ambient
conditions: as the pressure doesn't affect very much the value of
$c_\infty$ the agreement with the simulation data in the linear
region is extremely good. The open symbols, also reported in the
lower panel, indicate for some $Q$ values the position of the two
peaks found in the Transverse Current spectra, a quantity  that is
not experimentally observable, but that can be determined by MD
simulations. The position of these peaks clearly coincides with
the correspondent peak measured in the Longitudinal Current,
supporting the presence of a mixing phenomenon \cite{22}. We
further note that the energy range over which the weakly
dispersing feature is observed corresponds to the TA / TO branch
in hexagonal ice as determined by INS \cite{24} and IXS \cite{25}.

\begin{figure}[t]
\includegraphics[width=.55\textwidth]{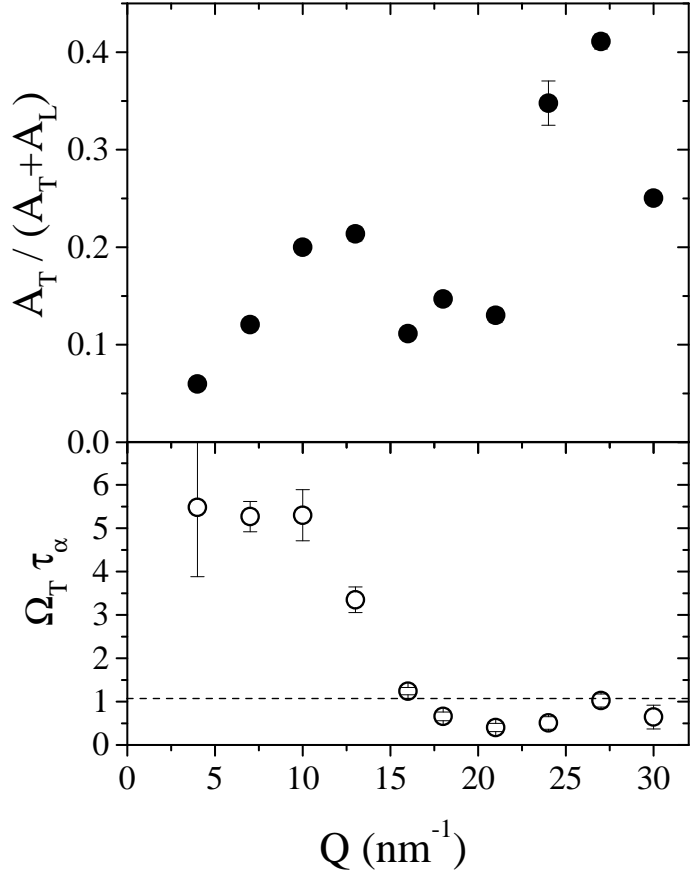}
\caption{\footnotesize{$Q$-dependence of the transverse to total
integrated intensity ratio (upper panel) and the product $\Omega_T
\tau_\alpha$ (lower panel) at T=263K. The dashed horizontal line
indicates the condition $\Omega_T \tau_\alpha$ = 1. }}
\label{AvsQ}
\end{figure}

To further characterize the second excitation we now study its
integrated intensity. Specifically, in figure \ref{AvsQ} we report
the "transverse" intensity $A_T$ normalized to the total
integrated intensity $A_T+A_L$ as a function of $Q$ for the
selected thermodynamic point $T$=263 K and $P$=2 kbar. In the
lower panel of Fig. \ref{AvsQ}, we report the evolution of
$\Omega_T \tau_\alpha$. We observe a general increasing trend of
the parameter $A_T/(A_T+A_L)$ with increasing $Q$. It is worth to
observe that the appearance of a transverse-like contribution with
an intensity increasing with $Q$ has also been registered in
experiments and simulations on glassy systems as glycerol
\cite{gly} and silica \cite{sio2a,sio2b,sio2c}. In silica the
mixing phenomenon turns out to be particularly enhanced both in MD
simulation and experiments indicating that, as in water, the local
tetrahedral structure favors the coupling of the L and T dynamics.
The intensity increase with $Q$ is expected on the basis of a
simple one-excitation picture within the harmonic approximation
for the dynamic structure factor, suggesting that, for a
non-dispersive excitation, $A_T/(A_T+A_L)\approx Q^2$. More
importantly, we also observe that, superimposed to the growth of
$A_T/(A_T+A_L)$ with $Q$, there is also a "modulation" in phase
with $\Omega_T \tau_\alpha$. When, at $Q\approx$ 15 nm$^{-1}$,
$\Omega_T \tau_\alpha$ become less than unity, the intensity of
the "transverse" peak decreases.

\begin{figure*}[t]
\hspace{-5cm}
\includegraphics[width=.75\textwidth]{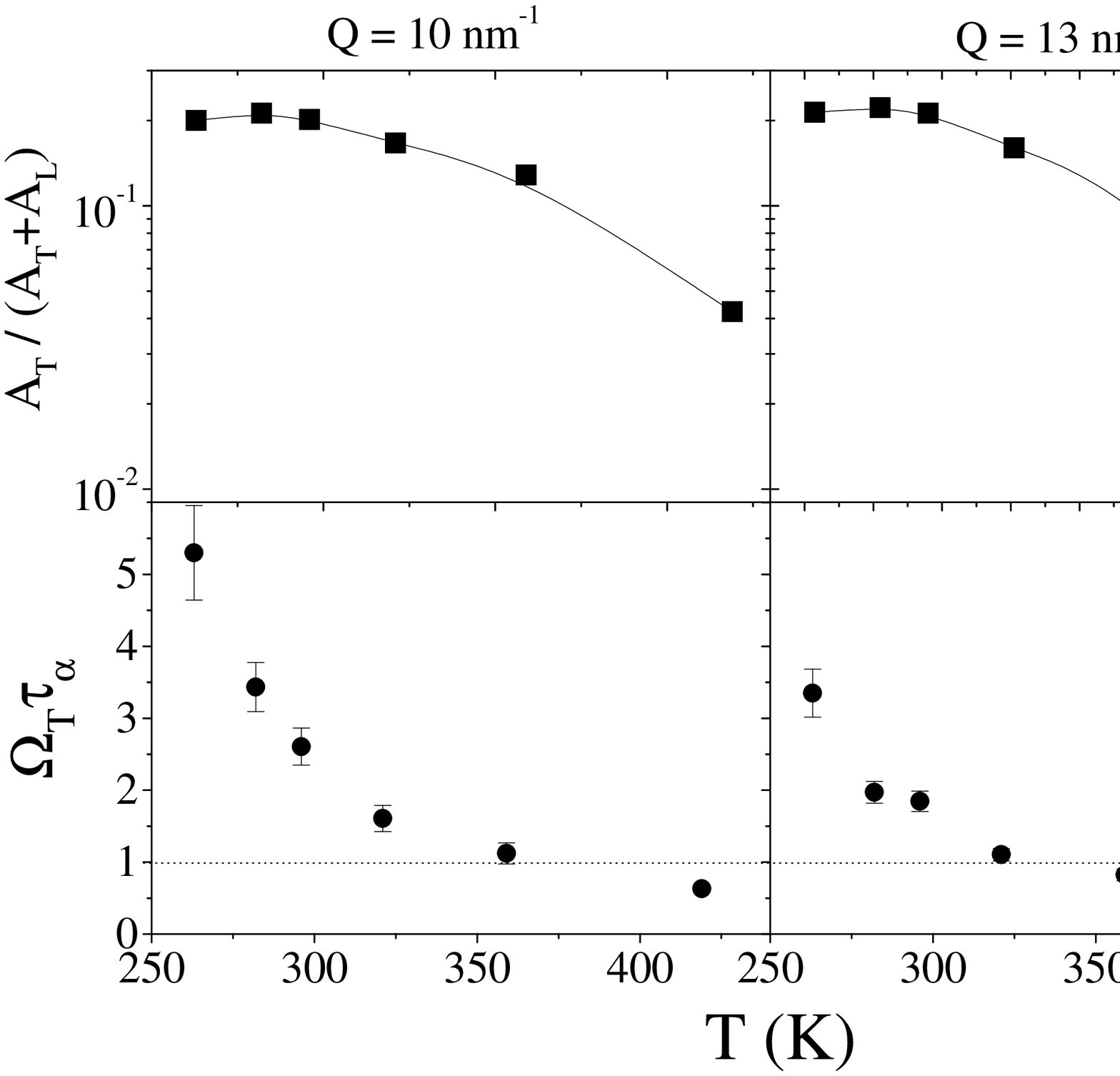}
\caption{\footnotesize{Top panels: temperature evolution of the
transverse to total integrated intensity ratio $ A_{T}/(A_T+A_L)$
at momentum transfers of 10 $ nm^{-1} $ (left panel) and 13 $
nm^{-1} $ (right panel). The solid lines are guides to the eye.
Bottom panels: temperature evolution of $\Omega_T \tau_\alpha$.
The dotted horizontal line marks the condition $\Omega_T
\tau_\alpha$ = 1.}} \label{AvsT}
\end{figure*}

The presence of a strong correlation between the intensity of the
transverse peak and the $\Omega_T \tau_\alpha$ is confirmed by
inspecting their temperature evolution reported in figure
\ref{AvsT} for two selected $Q$ values: $Q$ =10 and 13 nm$^{-1}$.
Here, when the time scale of the $\alpha$-relaxation is comparable
to the inverse of the frequency of the transverse excitation, the
ratio $A_T/(A_T+A_L)$ decreases, and for the highest temperature
becomes essentially zero. This behavior  -very different from that
of the longitudinal collective mode, which is affected by the
relaxation only in the sound velocity value- is just the one
expected for a transverse-like excitation: recovering a
liquid-like regime the system is no longer able to give an elastic
response to a shear stress and to sustain propagating transverse
waves. In this regime the transverse dynamics assumes a purely
relaxational behavior, corresponding to a peak at $\omega = 0$ in
the current spectrum \cite{17}. For this reason, still lacking a
formal model for the coupling, the fits at higher temperatures
have been performed just fixing the DHO parameters at the lowest T
values leaving the intensity $A_T$ free. This procedure allows us
to provide the information on the inelastic contribution of the
dispersion-less mode leaving the relaxational behavior accounted
for by the viscoelastic model. This procedure, as can be seen by
bare eye in the spectra at high T, doesn't affect at all the
statistical significance of the fits also when the transverse-like
contribution, which was mandatory at low T, can be completely
neglected.

\section{Conclusions}

In this paper the viscoelastic analysis of the IXS spectra of
water, already successfully performed in the low $Q$ range, has
been extended to much larger $Q$ values and in a wide range of
thermodynamical conditions.

At the high $Q$ values investigated here the presence of a second
excitation in the spectra cannot anymore be neglected and it has
been successfully taken into account in the analysis revealing a
behavior in agreement with the viscoelastic expectation for a
transverse-like excitation. This findings confirm our assignment
of a transverse nature to the dispersion-less mode of water
observed at an energy of $\approx$ 6 meV.

Furthermore the large variations of the structural relaxation time
(almost a decade) in the investigated thermodynamic range has
allowed us to perform an excellent test of the viscoelastic
picture. Summarizing, the two main results of the present work
are:
\begin{enumerate}
\item[(1)] The longitudinal-like dynamics can be properly
accounted for by a viscoelastic model also in the high $Q$ range:
indeed, besides the now well known transition from $c_O$ to
$c_\infty$ that takes place at small $Q$ ($Q \approx$ 4 nm$^{-1}$
at ambient conditions), for the first time the transition from
$c_\infty$ back to $c_0$, taking place at high Q values, could be
observed. This back-transition is a consequence of the De-Gennes
narrowing on the product $\Omega_L \tau_\alpha$ that becomes
smaller than one at $Q$ values around the $Q$'s of the main peak
in the $S(Q)$. \item[(2)] The second dispersion-less mode behaves
as a transverse-like excitation disappearing from the spectra as
soon as the structural relaxation process reaches the excitation
time scale ($\Omega_T \tau_\alpha$ = 1)
\end{enumerate}

\begin{figure}[h]
\vspace{0cm} \hspace{-2cm}
\includegraphics[width=.45\textwidth]{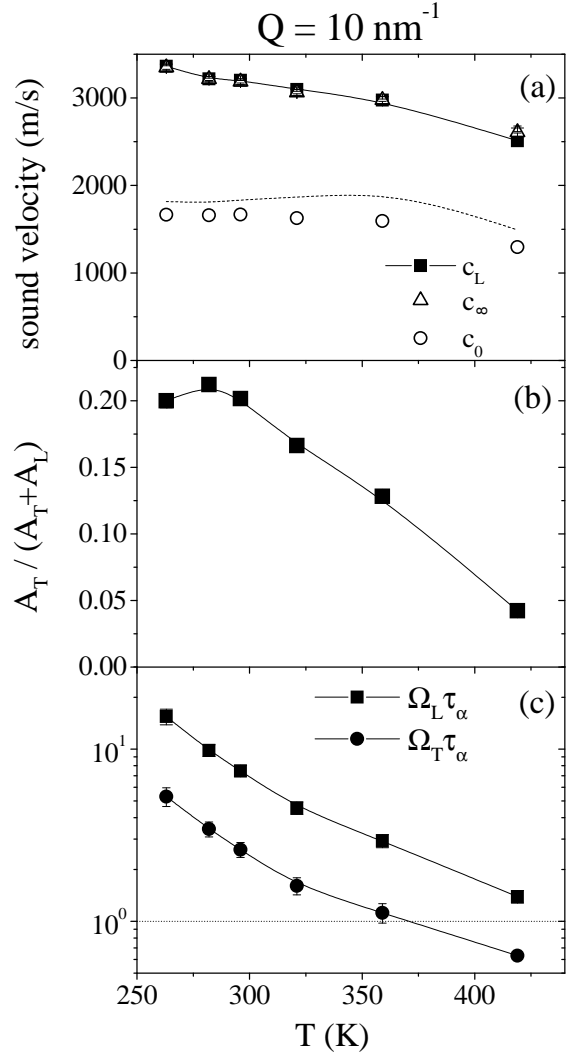}
\caption{\footnotesize{Comparison of the temperature evolution of
the L-like and T-like dynamics for a fixed $Q$ value ($Q$= 10 $
nm^{-1} $). (a) Evolution of the three velocities as in figure
\ref{svT}; the dashed line corresponds to the thermodynamic
prediction for $c_0$. (b) Temperature evolution of the transverse
to total integrated intensity ratio $A_{T}/(A_T+A_L)$. (c)
Temperature evolution of $\Omega_T \tau_\alpha$ and $\Omega_L
\tau_\alpha$; the dotted horizontal line marks the condition
$\Omega \tau_\alpha$ = 1 clearly showing that the two dynamics are
in different regimes at high $T$. The solid lines are guides to
the eye.}} \label{compLT}
\end{figure}

Finally, a few words must be devoted to the comparison between the
present -viscoelastic- explanation of the high frequency dynamics
of water and the alternative -solid based- description as
discussed in the introduction. To this purpose we compare in
figure \ref{compLT} the temperature behavior of the longitudinal
and transverse dynamics for a fixed $Q$ value ($Q$=10 nm$^{-1}$).
In panel (a) we report the velocities plot as in figure \ref{svT}
(here for a different $Q$ value) while in panel (b) we report the
(normalized) intensity of the dispersion-less mode at the same $Q$
value. The comparison of the two uppermost panels in figure
\ref{compLT} clearly demonstrates that while the apparent sound
velocity ($c_L$, full squares in panel (a)) is still far from from
$c_0$ (open dots) -i.~e. the system is in the elastic limit over
the whole $T$ range-, the intensity of the second excitation
becomes negligible at high $T$. Thus we have that, at high $T$ the
system is in the elastic regime, but the second excitation can no
longer be appreciated in the spectra. This evidence cannot be
framed in the interaction model proposed in Ref.~\cite{8,sacchnew}
that traces the "fast sound" phenomenon back to the presence and
the interaction between two modes. This observation, however, has
a clear explanation in the viscoelastic picture as illustrated in
panel (c). While the structural relaxation process has already
reached the transverse-like mode time scales, with $\Omega_T
\tau_\alpha <$ 1 for $T >$ 370 K thus entering the viscous regime,
it is still not affecting the higher energy longitudinal-like
excitation which still exhibits an elastic response ($\Omega_L
\tau_\alpha >$ 1 up to the highest explored temperature).

The data reported in Fig.~\ref{compLT}, together with the outcome
listed in items (1) and (2) before, are the main results of this
paper. All these results favor the viscoelastic description of the
high frequency dynamics of water. As a final observation, it must
be noted that the two models discussed before are not so different
as it appears at a first glance: in both cases it is the
interaction of the sound waves with another "mode" which causes
both the positive dispersion of the apparent sound velocity and
the appearance of a second peak in the dynamic structure factor.
In the viscoelastic case the second mode is purely relaxing, in
the solid-like approaches the second modes have a propagating
component. However, major differences arise in interpreting the
physical meaning of the parameters entering in the models: {\it
i)} in the solid-like model no information of the structural
relaxation time is contained, while $\tau_\alpha$ is directly
measurable in the viscoelastic model, it is in agreement with
independent determinations \cite{9c,10}, and its values allow to
calculate the viscosity, again in agreement with independent
measurements \cite{10}; {\it ii)} A further evidence, again in the
same direction, comes from the temperature dependence of the
energy of the second mode. In the solid-like framework, at
increasing temperature the data indicate a decrease of the
interaction parameter, corresponding to a decrease of the energy
position of the dispersion-less mode. This decrease of the energy
of the dispersion-less mode is not observed; for example, the
depolarized light scattering measurements \cite{sok} do not show
any noticeable shift in the position of the peak that corresponds
to the density of states of the second excitation; {\it iii)}
Finally, let us observe that the viscoelastic-based explanation of
the dynamics of the density fluctuations is actually considered
the proper approach in many liquids \cite{viscoOK} and it has been
substantiated by rigorous theories as the Mode Coupling Theory
\cite{MCT} (a theory that, via MD, has been proved to work
properly also in liquid water \cite{MCT_water}). In other words,
as the dynamic of liquid water shows the same phenomenology of
that of many other liquids, why, in water, should one look for a
different explanation?

\section{Acknowledgment}

We thank M.~A.~Ricci for pointing out the neutron diffraction
database \cite{ricci}.


\begin{references}

\bibitem{1} C.A. Angell, in {\itshape Water: A comprehensive Treatise},
 edited by F. Franks (Plenum, New York, 1981), Vol. 7.

\bibitem{2}P.G.~Debenedetti,{\itshape Metastable Liquids}
(Princeton University Press, 1996).

\bibitem{3}
C. Roenne, L. Thrane, P.-O. Astrand,
A. Wallqvist, K.V. Mikkelsen,
and S.R. Keiding, J. Chem. Phys. 107, 5319 (1997).

\bibitem{torre}
R. Torre, P. Bartolini, R. Righini, Nature 428, 296(2004)

\bibitem{4}L. van Hove, Phys. Rev. 95, 249 (1954).

\bibitem{6}
P. Bosi, F. Dupre', F. Menzinger, F. Sacchetti, and M. C.
Spinelli, Nuovo Cimento Lett. 21, 436 (1978).

\bibitem{5}
J. Teixeira, M.-C. Bellissent-Funel, S.H. Chen,
and B. Dorner, Phys. Rev. Lett. 54, 2681 (1985).

\bibitem{7}
F.J. Bermejo, M. Alvarez, S.M. Bennington, and
R. Vallauri, Phys. Rev. E 51, 2250 (1995).

\bibitem{8}C. Petrillo, F. Sacchetti, B. Dorner,
and J.-B. Suck, Phys. Rev. E 62, 3611 (2000).

\bibitem{sacchnew} F. Sacchetti, J.-B. Suck, C. Petrillo,
and B. Dorner, Phys. Rev. E 69, 061203 (2004).

\bibitem{9a} F. Sette, G. Ruocco, M. Krisch, U. Bergmann,
C. Masciovecchio, V. Mazzacurati, G. Signorelli, and R. Verbeni.
Phys. Rev. Lett. 75, 850 (1995).

\bibitem{9b} G. Ruocco, F. Sette, M. Krisch, U. Bergmann,
C. Masciovecchio, V. Mazzacurati, G. Signorelli, and R. Verbeni.
Nature 379, 521 (1996).

\bibitem{25} F. Sette, G. Ruocco, M. Krisch,
C. Masciovecchio, R. Verbeni, and U. Bergmann,
Phys. Rev. Lett. 77, 83 (1996).

\bibitem{9c} A. Cunsolo, G. Ruocco, F. Sette,
C. Masciovecchio, A. Mermet, G. Monaco, M. Sampoli,
and R. Verbeni.
Phys. Rev. Lett. 82, 775 (1999).

\bibitem{10} G. Monaco, A. Cunsolo, G. Ruocco,
and F, Sette, Phys. Rev. E 60, 5505 (1999).

\bibitem{9d}G. Ruocco and F. Sette,
J.Phys.: Cond. Matt. 11, R259 (1999).

\bibitem{11}M. Krisch et al.,
Phys. Rev. Lett. 89, 125502 (2002).

\bibitem{22}M. Sampoli, G. Ruocco, and
F. Sette, Phys. Rev. Lett. 79, 1678 (1997).

\bibitem{9e} It is worth to point out that this is not different
from what proposed in \cite{25} as the optic modes supported by
Petrillo et al. \cite{8} is the gap-less prosecution of the TA
mode in an extended BZ zone description of the phonon-like
dynamics.

\bibitem{14} EURISYS Mesures, Lingolsheim, France.

\bibitem{16} A. Saul and W. Wagner,
J. Phys. Chem. Ref. Data 18, 1537 (1989).

\bibitem{17} J.P. Boon and S. Yip,
{\itshape Molecular Hydrodynamics}
(Dover, New York, 1991).

\bibitem{18} U. Balucani and M. Zoppi,
{\itshape Dynamics of the Liquid State}
(MacGraw-Hill, New York, 1980).

\bibitem{n1} A. Cunsolo, G. Pratesi, R. Verbeni,
D. Colognesi, G. Monaco, C. Masciovecchio, G. Ruocco, and F.
Sette, J. Chem. Phys. {\bf 114}, 2259 (2001).

\bibitem{n2} T. Scopigno, F. Sette, G. Ruocco, and
G. Viliani, Phys. Rev. {\bf E66}, 031205 (2002).

\bibitem{n3} G. Monaco, D. Fioretto, L. Comez, and
G. Ruocco, Phys. Rev. {\bf E63}, 061502 (2001).

\bibitem{sio2b} G. Ruocco, and F. Sette,
J. of Phys.: Cond. Matt. {\bf 13}, 9141 (2001).

\bibitem{20}G. Harrison, {\itshape The Dynamical
Properties of Supercooled Liquids} (Academic Press,
New York, 1976).

\bibitem{masciov} C. Masciovecchio, S. C. Santucci,
A. Gessini, S. Di Fonzo, G. Ruocco, and F. Sette, Phys.
Rev. Lett. 92, 255507 (2004).

\bibitem{21}B. Fak and B. Dorner, Institut Laue Langevin
(Grenoble, France), technical report No. 92FA008S, (1992).

\bibitem{nonh} G. Ruocco, F. Sette, R. Di Leonardo,
G. Monaco, M. Sampoli, T. Scopigno, and G. Viliani, Phys. Rev.
Lett. 84, 5788 (2000).

\bibitem{ricci} http://www.isis.rl.ac.uk/disordered/Database/DBMain.htm

\bibitem{23} H.J.C. Berendsen, J.P.M. Postma, W.F. Van Gunsteren,
and H.J. Hermans, in {\itshape Intermolecular Forces}, edited by
B. Pulman (Reidel, Dordrecht, 1981), p. 331.

\bibitem{24}B. Renker, Phys. Lett. A 30, 493 (1969).

\bibitem{gly}
T. Scopigno, E. Pontecorvo, R. Di Leonardo, M. Krisch, G. Monaco,
G. Ruocco, B. Ruzicka, and F. Sette. Journal of Physics Condensed
Matter 15, S1269 (2003).

\bibitem{sio2a} R.~Dell'Anna, G.~Ruocco, M.~Sampoli,
and G.~Viliani, Phys. Rev. Lett. {\bf 80}, 1236 (1998).

\bibitem{sio2c} B. Ruzicka, T. Scopigno, S. Caponi,
A. Fontana, O. Pilla, P. Giura, G. Monaco, E. Pontecorvo,
G. Ruocco, F. Sette.
Physical Review B 69, 100201R (2004).

\bibitem{sok} A. P. Sokolov, J. Hurst and
D. Quitmann, Phys. Rev. B 51, 12865 (1995).

\bibitem{viscoOK} See for example: Journal of Physics: Condensed
Matter {\bf 15}, S737-1290, "Proceedings of the Third Workshop on
Non-equilibrium Phenomena in Supercooled Fluids, Glasses and
Amorphous Materials" (2003).

\bibitem{MCT} W.~Goetze, L.~Sjogren, Rep. Prog. Phys. {\bf 55},
241 (1992); W.~Goetze, J. Phys.: Cond. Matt. {\bf 11}, A1, (1999);
H.~.Z.~Cummins, J. Phys.: Cond. Matt. {\bf 11}, A95, (1999).

\bibitem{MCT_water} P.~Gallo, F.~Sciortino, P.~Tartaglia, and S.~H.~Chen,
Phys. Rev. Lett. {\bf 76}, 2730 (1996); F.~Sciortino, P.~Gallo,
P.~Tartaglia, and S.~H.~Chen, Phys Rev. E {\bf 54}, 6331 (1996);
S.~H.~Chen, P.~Gallo, F.~Sciortino, and P.~Tartaglia, Phys Rev. E
{\bf 56}, 4321 (1996); F.W. Starr, M.-C. Bellissent-Funel, and
H.~E.~Stanley, Phys. Rev. Lett. {\bf 82}, 3629 (1999);
F.~W.~Starr, F.~Sciortino, and H.~E.~Stanley, Phys. Rev. E {\bf
60}, 6757 (1999); L.~Fabbian, A.~Latz, R.~Schilling, F.~Sciortino,
P.~Tartaglia and C.~Theis, Phys. Rev. E {\bf 60}, 5768 (1999).

\end{references}
\end{document}